\documentstyle[preprint,epsfig,aps]{revtex}
\begin{document}
\draft
\preprint{\vbox{\hbox{IFT--P.012/98} \vspace{-0.3cm}
                \hbox{IFIC/98--22} \vspace{-0.3cm}
                \hbox{FTUV/98--22} \vspace{-0.3cm}
                \hbox{MADPH--98--1046}\vspace{-0.3cm}
                 \hbox{hep-ph/9802408}
 }}

\title{Bounds on Higgs and Gauge--Boson Interactions from LEP2 Data}

\author{O.\ J.\ P.\ \'Eboli $^{1,2}$, M.\ C.\ Gonzalez-Garcia $^{1,3}$, 
        S.\ M.\ Lietti $^{1}$ and S.\ F.\ Novaes $^{1}$}

\address{$^1$ Instituto de F\'{\i}sica Te\'orica, 
              Universidade Estadual Paulista, \\ 
              Rua Pamplona 145, 01405--900 S\~ao Paulo, Brazil.\\
         $^2$ Physics Department, University of Wisconsin, 
	      Madison, WI 53706, USA. \\
	 $^3$ Instituto de F\'{\i}sica Corpuscular IFIC/CSIC,
              Departament de F\'{\i}sica Te\`orica \\
              Universitat de Val\`encia, 46100 Burjassot,
	      Val\`encia, Spain.}
\date{February 20, 1998}
\maketitle
\begin{abstract}
\baselineskip 0.7 cm 
We derive bounds on Higgs and gauge--boson anomalous interactions
using the LEP2 data on the production of three photons and photon
pairs in association with hadrons. In the framework of $SU(2)_L
\otimes U(1)_Y$ effective Lagrangians, we examine all
dimension--six operators that lead to anomalous Higgs
interactions involving $\gamma$ and $Z$. The search for Higgs
boson decaying to $\gamma\gamma$ pairs allow us to obtain
constrains on these anomalous couplings that are comparable with
the ones originating from the analyses of $p\bar{p}$ collisions
at the Tevatron. Our results also show that if the coefficients
of  all ``blind'' operators are assumed to have same magnitude,
the indirect constraints on the anomalous couplings obtained from
this analyses, for Higgs masses  $M_H \lesssim$ 140 GeV, are more
restrictive than the ones coming from the $W^+W^-$ production.
\end{abstract}

\pacs{14.80.Cp, 13.85.Qk}

In the last few years it has been established that the
interactions of the gauge bosons with the fermions are well
described by the Standard Model (SM) \cite{lepew}. However, we
are just beginning to directly probe the self--interactions of
the electroweak gauge bosons through their pair production at the
Tevatron \cite{tev} and LEP2 \cite{lepii} colliders. 

On the other hand, we still do not have any experimental evidence
on how the symmetry breaking takes place in the SM. A larger
symmetry breaking sector can introduce modifications in the
interactions of the vector and Higgs bosons predicted by the SM.
These possible deviations of the gauge--boson couplings from
their SM values can be parametrized through the use of effective
Lagrangians.  When the $SU(2)_L \otimes U(1)_Y$ symmetry is
realized linearly in the effective theory, {\em i.e.\/} when
there is a light scalar Higgs doublet in the spectrum, the lowest
order anomalous interactions are given by dimension--six
operators \cite{buch}. These new interactions can alter
considerably the low energy phenomenology. For instance, some
operators can give rise to anomalous $H\gamma\gamma$ and
$HZ\gamma$ couplings which may affect the Higgs boson production
and decay \cite{Hagiwara2}. 

It is important to notice that, since the linearly realized
effective Lagrangians relate the modifications in the Higgs
couplings to the ones in the vector boson vertex
\cite{buch,Hagiwara2,linear,hisz,dim6:zep}, the search for Higgs
bosons can be used to not only study its properties, but also to
place bounds on the gauge--boson self interactions. This approach
is more efficient when the analyses is performed for decays of
the Higgs boson that are suppressed in the SM, such as $H
\rightarrow \gamma\gamma$ that occurs only at one loop level, and
are enhanced by new anomalous interactions \cite{ee,fromtev}.

In this work, we use the recently released LEP data on the
production of $\gamma\gamma$ in association with hadrons
\cite{opal:gg} and $\gamma\gamma\gamma$ \cite{opal:ggg} to
constrain possible Higgs--boson anomalous couplings to
vector--bosons. Working with effective operators linearly
invariant under the $SU(2)_L \otimes U(1)_Y$, we obtain indirect
limits on anomalous gauge--boson interactions from the search of
Higgs bosons decaying into two photons. Our results show that,
for Higgs masses $M_H \lesssim$ 140 GeV, the constraints on
anomalous couplings obtained from this analyses are more
restrictive than the ones coming from the $W^+W^-$ production.

In the linear representation of the $SU(2)_L \otimes U(1)_Y$
symmetry breaking mechanism, the SM model is the lowest order
approximation while the first corrections, which are of dimension
six, can be written as
\begin{equation}
{\cal L}_{\text{eff}} = \sum_n \frac{f_n}{\Lambda^2} {\cal O}_n
\label{l:eff}
\end{equation}
where the operators ${\cal O}_n$ involve vector--boson and/or
Higgs--boson fields with couplings $f_n$. This effective
Lagrangian describes well the phenomenology of models that are
somehow close to the SM since a light Higgs scalar doublet is
still present at low energies. Of the eleven possible operators
${\cal O}_{n}$ that are $P$ and $C$ even, only six of them modify
the Higgs--boson couplings to vector bosons
\cite{Hagiwara2,dim6:zep},
\begin{eqnarray}
&&{\cal O}_{BW} =  \Phi^{\dagger} \hat{B}_{\mu \nu} 
\hat{W}^{\mu \nu} \Phi \; , \nonumber \\ 
&&{\cal O}_{WW} = \Phi^{\dagger} \hat{W}_{\mu \nu} 
\hat{W}^{\mu \nu} \Phi  \; , \nonumber \\
&&{\cal O}_{BB} = \Phi^{\dagger} \hat{B}_{\mu \nu} 
\hat{B}^{\mu \nu} \Phi \; ,  
\label{eff}  \\
&&{\cal O}_W  = (D_{\mu} \Phi)^{\dagger} 
\hat{W}^{\mu \nu}  (D_{\nu} \Phi) \; , \nonumber \\
&&{\cal O}_B  =  (D_{\mu} \Phi)^{\dagger} 
\hat{B}^{\mu \nu}  (D_{\nu} \Phi)  \; , \nonumber \\
&&{\cal O}_{\Phi,1} = \left ( D_\mu \Phi \right)^\dagger \Phi^\dagger \Phi
\left ( D^\mu \Phi \right ) \; , \nonumber
\end{eqnarray}
where $\Phi$ is the Higgs doublet, $D_\mu$ the covariant
derivative, $\hat{B}_{\mu \nu} = i (g'/2) B_{\mu \nu}$, and
$\hat{W}_{\mu \nu} = i (g/2) \sigma^a W^a_{\mu \nu}$, with
$B_{\mu \nu}$ and $ W^a_{\mu \nu}$ being respectively the
$U(1)_Y$ and $SU(2)_L$ field strength tensors.

Anomalous $H\gamma\gamma$, $HZ\gamma$, and $HZZ$ couplings are
generated by (\ref{eff}), which, in the unitary gauge, are given
by
\begin{eqnarray}
{\cal L}_{\text{eff}}^{\text{H}} &=& 
g_{H \gamma \gamma} H A_{\mu \nu} A^{\mu \nu} + 
g^{(1)}_{H Z \gamma} A_{\mu \nu} Z^{\mu} \partial^{\nu} H + 
g^{(2)}_{H Z \gamma} H A_{\mu \nu} Z^{\mu \nu}
\nonumber \\
&+& g^{(1)}_{H Z Z} Z_{\mu \nu} Z^{\mu} \partial^{\nu} H + 
g^{(2)}_{H Z Z} H Z_{\mu \nu} Z^{\mu \nu} +
h^{(3)}_{H Z Z} H Z_\mu Z^\mu
\; , 
\label{H} 
\end{eqnarray}
where $A(Z)_{\mu \nu} = \partial_\mu A(Z)_\nu - \partial_\nu
A(Z)_\mu$. The effective couplings $g_{H \gamma \gamma}$,
$g^{(1,2)}_{H Z \gamma}$, and $g^{(1,2,3)}_{H Z Z}$ are related
to the coefficients of the operators appearing in (\ref{l:eff})
through,
\begin{eqnarray}
g_{H \gamma \gamma} &=& - \left( \frac{g M_W}{\Lambda^2} \right)
                       \frac{s^2 (f_{BB} + f_{WW} - f_{BW})}{2} \; , 
\nonumber \\
g^{(1)}_{H Z \gamma} &=& \left( \frac{g M_W}{\Lambda^2} \right) 
                     \frac{s (f_W - f_B) }{2 c} \; ,  
\nonumber \\
g^{(2)}_{H Z \gamma} &=& \left( \frac{g M_W}{\Lambda^2} \right) 
                      \frac{s [2 s^2 f_{BB} - 2 c^2 f_{WW} + 
                     (c^2-s^2)f_{BW} ]}{2 c}  \; , 
\label{g} \\ 
g^{(1)}_{H Z Z} &=& \left( \frac{g M_W}{\Lambda^2} \right) 
	              \frac{c^2 f_W + s^2 f_B}{2 c^2} \nonumber \; , \\
g^{(2)}_{H Z Z} &=& - \left( \frac{g M_W}{\Lambda^2} \right) 
  \frac{s^4 f_{BB} +c^4 f_{WW} + c^2 s^2 f_{BW}}{2 c^2} \nonumber \; , \\
g^{(3)}_{H Z Z} &=& 2 \left( \frac{ M_W^3}{g \Lambda^2} \right) 
	              \frac{f_{\Phi,1}}{c^2} \; , \nonumber
\end{eqnarray}
with $g$ being the electroweak coupling constant, and $s(c)
\equiv \sin(\cos)\theta_W$. 

The operators ${\cal O}_{\Phi,1}$ and ${\cal O}_{BW}$ contribute
at tree level to the vector--boson two--point functions, and
consequently are severely constrained by the low--energy data
\cite{dim6:zep}. The present limits on these operators for
$M_H=100$ GeV and $m_{\text{top}}=175$ GeV read \cite{rob},
\begin{equation}
\frac{f_{\Phi,1}}{\Lambda^2}  = (0.3 \pm 0.16) \hbox{ TeV}^{-2}
\;\; , \;\;\;\;
\frac{f_{BW}}{\Lambda^2}  = (3.7 \pm 2.4) \hbox{ TeV}^{-2} \; .
\label{phi:bw}
\end{equation}
Consequently we will neglect these operators in our analyses.  On
the order hand, the remaining operators are indirectly
constrained via their one--loop contributions to low energy
observables, which leads to $f_i/\Lambda^2 \sim 100$
TeV$^{-2}$. The present data on gauge--boson pair production
leads to the following 95\% CL bounds on anomalous couplings
\cite{tev,lepii},
\begin{equation}
\left | \frac{f_W}{\Lambda^2} \right | < 300 \mbox{ TeV}^{-2} 
\;\; , \;\;\;\;
\left | \frac{f_B}{\Lambda^2} \right | < 390 \mbox{ TeV}^{-2} \; .
\end{equation}

\begin{figure}
\begin{center}
\mbox{\epsfig{file=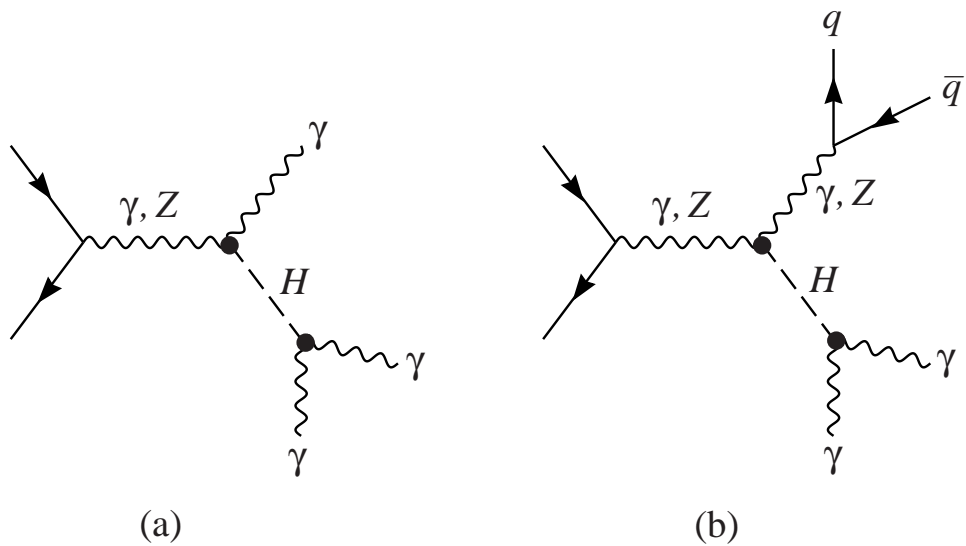,width=0.8\textwidth}}
\end{center}
\caption{Anomalous contribution for the $\gamma\gamma\gamma$
production (a) and $\gamma\gamma$ in association with hadrons
(b).}
\label{fig:feyn}
\end{figure}

In order to obtain constraints on the anomalous couplings
described above, we have used the recent OPAL
data \cite{opal:gg,opal:ggg} for the reactions,
\begin{eqnarray}
e^+ e^- &\rightarrow& \gamma \gamma \gamma \; ,
\label{ggg}
\\
e^+ e^- &\rightarrow& \gamma \gamma + \mbox{ hadrons} \; .
\label{ggh}
\end{eqnarray}
The Feynman diagrams describing the anomalous contributions to
the above reactions are displayed in Fig.\ \ref{fig:feyn}.  The
scattering amplitudes were generated using Madgraph \cite{madg}
and Helas \cite{hellas}, with the anomalous couplings, arising
from the operators (\ref{eff}), being implemented as Fortran
routines. In Refs.\ \cite{opal:gg,opal:ggg}, data taken at
several energy points in the range $\sqrt{s}=130 \,(91)$--$172$,
for the $\gamma\gamma\gamma$ ($\gamma \gamma + \mbox{ hadrons}$)
are combined. In our calculation we also combined the expected
number of events for the corresponding energies and accumulated
luminosities.

It is important to notice that the dimension-six operators
(\ref{eff}) do not induce $4$--point anomalous couplings like $Z
Z \gamma \gamma$, $Z \gamma\gamma \gamma$, and $\gamma \gamma
\gamma \gamma$, being these terms generated only by
dimension--eight and higher operators. Since the production and
decay of the Higgs boson also involve two dimension--six
operators, we should, in principle, include in our calculations
dimension--eight operators that contribute to the above
processes. Notwithstanding, we can neglect the higher order
interactions and bound the dimension--six couplings under the
naturalness assumption that no cancelation takes place amongst
the dimension--six and --eight contributions that appear at the
same order in the expansion. 

\begin{figure}
\begin{center}
\mbox{\epsfig{file=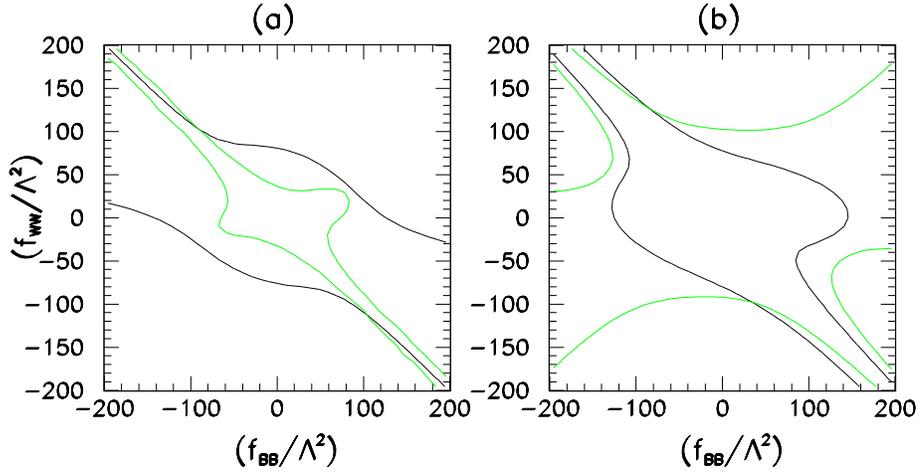,width=0.8\textwidth}}
\end{center}
\vskip -5 cm
\caption{Contour plot of $f_{BB} \times f_{WW}$, in TeV$^{-2}$.
The curves show the 95\% CL deviations from the SM total cross
section, for $e^+ e^- \to \gamma \gamma \gamma$ (dark lines) and
$e^+ e^- \to q \bar{q} \gamma \gamma$ (light lines) for (a)
$M_H = 60$ GeV and (b) $M_H = 80$ GeV. The excluded regions are
outside the lines}
\label{fig:ggg}
\end{figure}

\begin{figure}
\begin{center}
\mbox{\epsfig{file=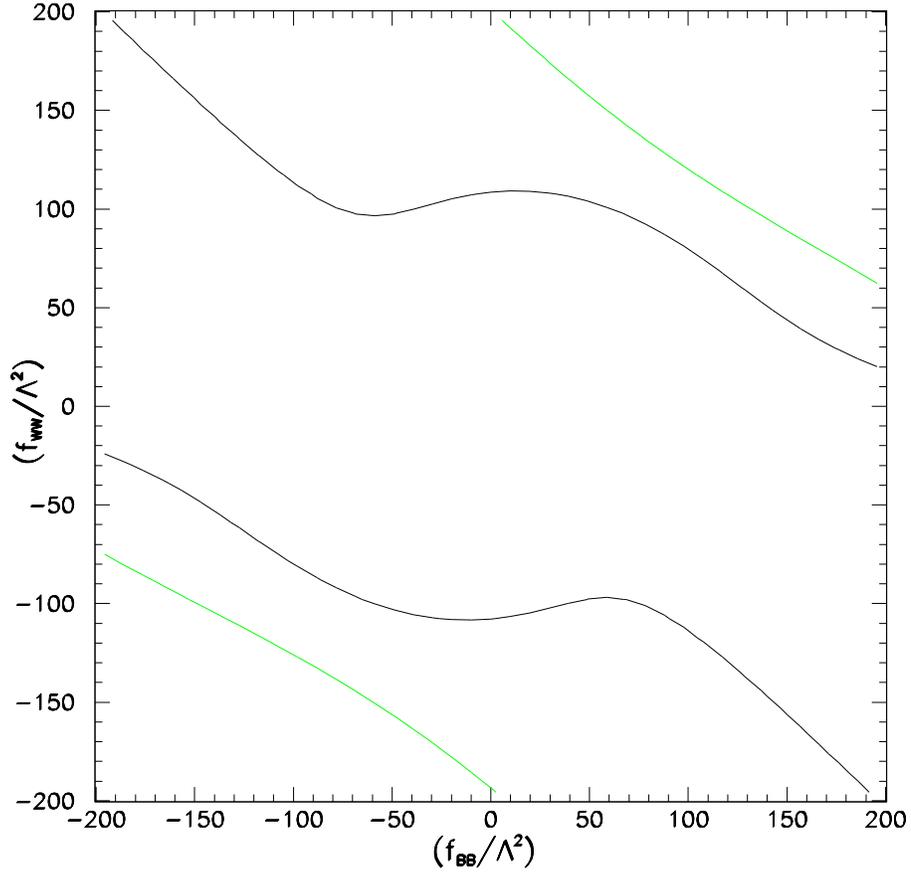,width=0.8\textwidth}}
\end{center}
\caption{Contour plot of $f_{BB} \times f_{WW}$, in TeV$^{-2}$.
The curves show the 95\% CL deviations from the SM total cross
section, for $e^+ e^- \to \gamma \gamma \gamma$ with $M_H = 100$
GeV (dark lines), and $M_H = 120$ GeV (light lines).}
\label{fig:ggh}
\end{figure}

\begin{figure}
\begin{center}
\mbox{\epsfig{file=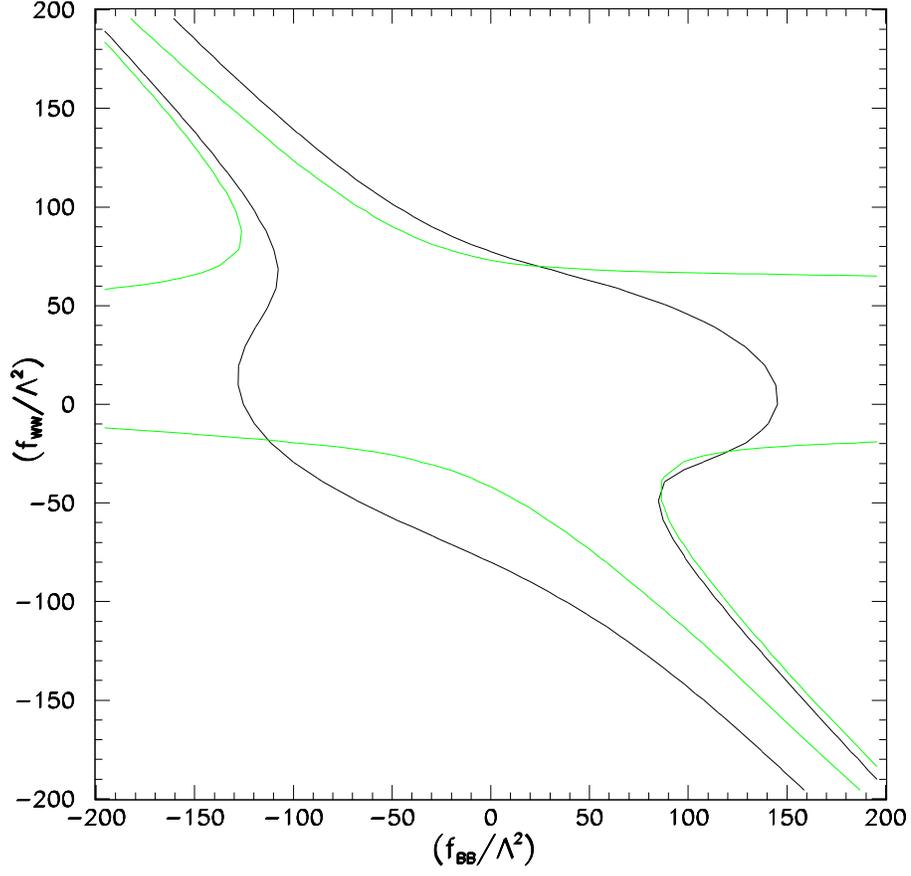,width=0.8\textwidth}}
\end{center}
\caption{Contour plot of $f_{BB} \times f_{WW}$, in TeV$^{-2}$.
The curves show the 95\% CL deviations from the SM total cross
section, for $e^+ e^- \to \gamma \gamma \gamma$ with $M_H = 80$
GeV (dark lines) and $p \bar{p} \to \gamma \gamma + \not \! E_T$
at Tevatron with $M_H = 80$ GeV (light lines).}
\label{fig:tev}
\end{figure}

We start our analyses assuming that the only non--zero
coefficients are the ones that generate the anomalous
$H\gamma\gamma$, {\it i.e.}, $f_{BB}$ and $f_{WW}$.  We exhibit
in Figs.\ \ref{fig:ggg} and \ref{fig:ggh} the 95\% CL exclusion
region in the plane $f_{BB} \times f_{WW}$ obtained from the OPAL
data on multiple photon production \cite{opal:ggg} and diphoton
events exhibiting hadrons \cite{opal:gg}. In this analyses we set
all other anomalous couplings to zero and evaluated only the
anomalous contribution as the SM backgrounds were already
subtracted in the experimental results. For small Higgs masses
(see Fig.\ \ref{fig:ggg}) the $Z$, which decays hadronically, can
be produced on mass shell and, therefore, the strongest bounds
come from the diphoton production in association with hadrons.
Since the anomalous contribution to $H\gamma\gamma$ is zero for
$f_{BB} = - f_{WW}$, the bounds become very weak close to this
line, as is clearly shown in Fig.\ \ref{fig:ggg}.  For higher
Higgs--boson masses ($M_H \gtrsim 80$ GeV), the $Z$ cannot be
on--mass shell, and the $\gamma\gamma$ production accompanied by
hadrons is suppressed. In this case, only the
$\gamma\gamma\gamma$ final state is able to lead to new bounds.
Moreover, the anomalous production of a $H\gamma$ pair is also
suppressed by the phase space as $M_H$ increases and the limits
worsen, as we can see from Fig.\ \ref{fig:ggh}. It is interesting
to notice that the bounds obtained using the above processes are
of the same order of the ones that can be extracted from the
Tevatron collider for small Higgs boson masses ($M_H \lesssim 80$
GeV). For the sake of comparison, we present in Fig.\
\ref{fig:tev} the contours in the $f_{BB} \times f_{WW}$ plane
from analyses of $e^+ e^- \to \gamma\gamma\gamma$ and from $p
\bar{p} \rightarrow H (\rightarrow \gamma\gamma) + \not \! E_T$
\cite{fromtev}. Therefore, LEP2 should lead to more stringent
bounds on dimension--six operators with the increase of its
accumulated luminosity.

In order to reduce the number of free parameters and, at the same
time, relate the anomalous Higgs and the triple vector--boson
couplings, one can make the assumption that all blind operators
affecting the Higgs interactions have a common coupling $f$, {\it
i.e.}
\begin{equation}
f_W = f_B = f_{WW} = f_{BB} = f \; , 
\label{blind}
\end{equation}
and that $f_{\Phi,1} \simeq f_{BW} \simeq 0$
\cite{Hagiwara2,dim6:zep,dpf}. In this scenario, $g^{(1)}_{H Z
\gamma} = g^{(3)}_{HZZ} = 0$, and we can relate the Higgs boson
anomalous coupling $f$ with the conventional parametrization of
the vertex $WWV$ ($V=Z$, $\gamma$) \cite{hhpz}
\begin{equation}
\Delta \kappa_\gamma = \frac{M_W^2}{\Lambda^2}~ f
\;\; , \;\;\;\;
\Delta \kappa_Z = \frac{M^2_Z}{2 \Lambda^2}~ ( 1 - 2 s^2)~ f 
\;\; , \;\;\;\;
\Delta g^Z_1 = \frac{M^2_Z}{2 \Lambda^2}~ f \; .
\label{trad} 
\end{equation}

We present in Table \ref{tab:f} the 95\% CL allowed regions of
the anomalous couplings in the scenario defined by Eq.\
(\ref{blind}). In this framework, the bounds become weaker with
the increase of the Higgs boson mass. The production of diphotons
in association with hadrons is again important only when its is
possible to produce a pair $HZ$ on mass shell. Using the
relations (\ref{trad}), it is possible to translate these bounds
into limits on triple gauge bosons couplings $\Delta
\kappa_\gamma$, $\Delta \kappa_Z$, and $\Delta g^Z_1$, which we
show in Table \ref{tab:kappa} for the $\gamma\gamma\gamma$
production. As can be seen from this Table, the search for Higgs
bosons decaying into photon pairs leads to limits substantially
better then the ones derived from the recent analyses of $W^+W^-$
production at LEP2 \cite{lepii}.

Summarizing, in this work we have estimated the limits on anomalous
dimension--six Higgs boson interactions that can be derived from
the existing data on the search for Higgs bosons decaying into
two photons at LEP2. The bounds that arise from the anomalous
Higgs boson searches at LEP2 are as restrictive as the ones
obtained at the Tevatron for small Higgs masses ($M_H \lesssim
80$ GeV). Under the assumption of equal coefficients for all
anomalous Higgs operators, these bounds also lead to limits on
triple--gauge--boson couplings. Our results show that the limits
obtained through this search are more restrictive than the ones
derived from the $W$ pair production analyses. 

\acknowledgments
We would like to thank Kirsten Sachs and Peter Maettig from
valuable discussions.  M.\ C.\ G--G is grateful to the Instituto
de F\'{\i}sica Te\'orica for its kind hospitality.  O.\ J.\ P.\
E.\ is grateful to the Physics Department of University of
Wisconsin, Madison for its kind hospitality.  This work was
supported by Funda\c{c}\~ao de Amparo \`a Pesquisa do Estado de
S\~ao Paulo (FAPESP), by DGICYT under grant PB95-1077, by CICYT
under grant AEN96--1718, and by Conselho Nacional de
Desenvolvimento Cient\'{\i}fico e Tecnol\'ogico (CNPq), by the
University of Wisconsin Research Committee with funds granted by
the Wisconsin Alumni Research Foundation, and by the U.S.\
Department of Energy under Grant No.~DE-FG02-95ER40896.



\begin{table}
\begin{tabular}{||c||c||c||}
$M_H$(GeV) & $e^+ e^- \to \gamma \gamma \gamma$ & 
$e^+ e^- \to q \bar{q} \gamma \gamma$ \\
\hline 
\hline
60 & (  $-$56  ,   50  ) & ( $-$24  , 35  ) \\
\hline
80 & (  $-$53  ,   49  ) & ( $-$107  , 128  ) \\ 
\hline
100 & ( $-$64  ,   57  ) & ( $-$730  , 750  )\\ 
\hline
120 & ( $-$82  ,   70  ) & ------ \\ 
\hline
140 & ( $-$192  , 175  ) & ------ 
\end{tabular}
\medskip
\caption{Allowed range of $f/\Lambda^2$ in TeV$^{-2}$ at 95\%
CL coming from the  processes $e^+ e^- \to \gamma \gamma
\gamma$ and $e^+ e^- \to q \bar{q} \gamma \gamma$ at LEP2.
We  assumed the scenario defined by Eq.\ (\protect\ref{blind}).}
\label{tab:f}
\end{table}

\begin{table}
\begin{tabular}{||c||c||c||c||}
$M_H$(GeV) & $\Delta \kappa_\gamma$ & $\Delta \kappa_Z$ & $\Delta g^Z_1$ \\
\hline 
\hline
60 & ( $-$0.36  , 0.32  ) & ( $-$0.13 , 0.11) & ($-$0.23 ,  0.21) \\
\hline
80 & ( $-$0.34  , 0.32  ) & ($-$0.12  ,  0.11) & ($-$0.22  ,  0.21) \\
\hline
100 & ( $-$0.41  , 0.37  ) & ($-$0.15 ,  0.13) & ($-$0.26  ,  0.24) \\
\hline
120 & ( $-$0.53  , 0.45  ) & ($-$0.19 ,  0.16) & ($-$0.34  ,  0.29) \\
\hline
140 & ( $-$1.24  , 1.13  ) & ($-$0.44 ,  0.40) & ($-$0.80  ,  0.73) 
\end{tabular}
\medskip
\caption{95\%  CL allowed range of $\Delta\kappa_\gamma$, $\Delta
\kappa_z$, and $\Delta g^Z_1$ obtained from the analyses of
$\gamma\gamma\gamma$ production,  assuming the scenario defined
by Eq.\ (\protect\ref{blind}).}
\label{tab:kappa}
\end{table}

\end{document}